\documentclass[9pt,conference]{IEEEtran}
\IEEEoverridecommandlockouts

\usepackage{cite}
\usepackage{amsmath,amssymb,amsfonts}
\usepackage{algorithmic}
\usepackage{graphicx}
\usepackage{textcomp}
\usepackage{booktabs}
\usepackage{url}
\def\BibTeX{{\rm B\kern-.05em{\sc i\kern-.025em b}\kern-.08em
    T\kern-.1667em\lower.7ex\hbox{E}\kern-.125emX}}
\begin{document}

\title{The SonicAGI System for the REAL-TSE Challenge}

\author{\IEEEauthorblockN{Kai Li\textsuperscript{1,2,$\dagger$}, Wendi Sang\textsuperscript{1,$\dagger$}, Jintao Cheng\textsuperscript{1}, Xiaolin Hu\textsuperscript{1,2,3,$\ast$}}
\IEEEauthorblockA{
\textsuperscript{1}Department of Computer Science and Technology, Institute for AI, BNRist, Tsinghua University, Beijing, China\\
\textsuperscript{2}IDG/McGovern Institute for Brain Research, Tsinghua University, Beijing, China\\
\textsuperscript{3}Chinese Institute for Brain Research (CIBR), Beijing, China\\
\textsuperscript{$\dagger$}These authors contributed equally to this work. \textsuperscript{$\ast$}Corresponding author.}
}

\maketitle

\begin{abstract}
Real-world target speaker extraction (TSE) remains challenging because target speech, interference, and enrollment are recorded under mismatched acoustic conditions with reverberation, noise, and irregular conversational overlap. This paper describes the SonicAGI submission to the REAL-TSE Challenge (IEEE SLT 2026). We take a data-centric approach that combines fully simulated mixtures from clean speech with real meeting overlaps, and use a frozen offline enhancer to provide a denoised mirror of real targets for auxiliary supervision. For the online track, we introduce SwiftNet-Lookahead, which inserts a single bounded-lookahead module before a strictly causal iterative separator and keeps the total system latency at 96 ms. For the offline track, we use a frame-level enrollment cross-attention USEF-TFGridNet with a magnitude-domain fusion stage designed to balance perceptual quality and speaker fidelity. In the final verified Phase-1 ranking, SwiftNet-Lookahead ranks third in Track~1 and USEF-TFGridNet ranks fifth in Track~2, both exceeding the challenge baselines. These results suggest that real-data-oriented training and track-specific modeling are effective for conversational TSE.
\end{abstract}


\section{Introduction}
Target speaker extraction (TSE) extracts a designated speaker from a mixture using an enrollment utterance \cite{wang2024overview}. It is a key front-end for hearing aids, meeting transcription, and speech interaction. Although recent speech and query-based separation methods perform well on curated or synthetic mixtures \cite{li2025advances,li2025efficient,li2026semantically}, real conversational recordings contain reverberation, background noise, irregular overlap, and cross-scene enrollment. These factors expose a persistent mismatch between synthetic training and real evaluation conditions, making data realism a central bottleneck.

The REAL-TSE Challenge \cite{realtseChallenge2026} evaluates this setting on real Mandarin and English meeting and dinner-party recordings, scored by transcription error, target-speaker timing, speaker similarity, and perceptual quality. Track~1 requires online processing below 100 ms algorithmic latency and verifies causality through input-perturbation tests, whereas Track~2 allows full-context offline processing. Development data are held-out partitions of AISHELL-4 \cite{fu2021aishell4}, AliMeeting \cite{yu2022m2met}, AMI \cite{carletta2006ami}, DipCo \cite{vansegbroeck2020dipco}, and CHiME-6 \cite{watanabe2020chime6}; their development and test splits are excluded from training.

Our SonicAGI submission is built around three components. First, we construct training data with a two-route sampler: \emph{Fully Simulated Mixing} provides controlled speaker overlap from clean speech, while \emph{Real-recording Mixing} samples overlapping utterances from real meetings; a frozen offline enhancer further produces a cleaned mirror of real targets for auxiliary supervision. Second, for Track~1, we propose \emph{SwiftNet-Lookahead}, which inserts a single chunked bidirectional LSTM lookahead module before a strictly causal Swift-Net separator, so that future context is available within the latency budget but does not accumulate across iterative updates. Third, for Track~2, we use a frame-level enrollment cross-attention USEF-TFGridNet and apply magnitude-domain fusion between the extractor output and its enhanced version, designed to balance perceptual quality and speaker fidelity. The resulting systems achieve third place in Track~1 and fifth place in Track~2, both outperforming the challenge baselines.

\section{Data Construction}
All audio is 16 kHz mono, and training samples are generated on the fly. Each sample takes, with equal probability, one of two routes: a \emph{Fully Simulated Mixing} route that builds the mixture from clean speech under simulated acoustic conditions, and a \emph{Real-recording Mixing} route that remixes target and interferer segments cut from real meetings. The two routes share the enrollment-length sampling and the global-gain setting; enrollment lengths are drawn from $[3,15]$ s (covering 94\% of the development-set durations) and the training chunk is 3 s. Table~\ref{tab:datacfg} lists the main simulation parameters.

\begin{table}[htbp]
\caption{Key data-simulation parameters (shared by both routes unless otherwise noted).}
\begin{center}
\footnotesize
\setlength{\tabcolsep}{3pt}
\begin{tabular}{@{}ll@{}}
\toprule
\textbf{Parameter} & \textbf{Value} \\
\midrule
Sample rate & 16 kHz, mono \\
Fully Simulated / Real-recording Mixing prob. & 0.5 / 0.5 \\
Training chunk length & 3.0 s \\
Enrollment length & $[3, 15]$ s \\
Number of speakers $N$ & $2$--$4$ \\
Signal-to-interference ratio & $[-5, 10]$ dB \\
Noise probability (synthetic) & 0.6 \\
Noise SNR (synthetic) & $[-5, 25]$ dB \\
Reverberation RT60 (synthetic) & $[0.1, 0.7]$ s \\
Global gain & $[-12, 0]$ dB \\
Cross-scene enrollment prob.\ (real) & 0.5 \\
Samples per epoch & 20{,}000 \\
\bottomrule
\end{tabular}
\label{tab:datacfg}
\end{center}
\end{table}

\subsection{Source corpora}
The Fully Simulated Mixing route draws clean single-speaker speech from LRS3 \cite{afouras2018lrs3} and VoxCeleb2 \cite{chung2018voxceleb2}, transcoded from mp4 to PCM WAV. The Real-recording Mixing route draws overlapping speech from the official \emph{training} splits of AISHELL-4 \cite{fu2021aishell4}, AliMeeting \cite{yu2022m2met}, AMI \cite{carletta2006ami} (single distant microphone), and CHiME-6 \cite{watanabe2020chime6}. Following the challenge rules, we exclude the development and test sets of every corpus, and never use DipCo for training. WHAM! \cite{wichern2019wham} noise serves as the additive background in the Fully Simulated Mixing route.

\subsection{Fully Simulated Mixing route}
A Fully Simulated Mixing sample first selects a target speaker and an anchor utterance, then draws $N\in[2,4]$ speakers in total (of which $N-1$ interferers come from the full speaker pool), and reverberates each source with an independent single-channel room impulse response generated by a fast random-image simulator \cite{luo2022frarir} (RT60 in $[0.1,0.7]$ s). Interferers are added at a signal-to-interference ratio in $[-5,10]$ dB; the mixture is peak-normalized jointly with the sources, an optional global gain in $[-12,0]$ dB is applied, and finally WHAM! noise is added with probability $0.6$ at $[-5,25]$ dB. The enrollment uses \emph{another} utterance of the target speaker, and is degraded \emph{independently} of the mixture with its own room impulse response (probability $0.5$) and additive noise (probability $0.5$, $[0,20]$ dB), so that the model cannot use a shared acoustic channel as a shortcut cue for identifying the target.

\subsection{Real-recording Mixing route}
A Real-recording Mixing sample is drawn entirely from one meeting: the target and one or more interferers are real utterances from the same recording, so the mixture keeps genuine acoustics and we add no artificial reverberation or noise \cite{li2025sonicsim}. Interferers are remixed at a signal-to-interference ratio in $[-5,10]$ dB, followed by the same peak normalization and optional global gain. With probability $0.5$ the enrollment prefers a \emph{cross-scene} reference (the same speaker in another meeting), falling back to a same-scene one otherwise; this suppresses the model's shortcut dependence on the acoustic environment and matches the cross-meeting enrollment of the evaluation set.

Real meeting segments are never perfectly clean: the ``target'' still contains background noise and crosstalk. To use these segments without forcing the network to reproduce that noise, we pre-compute a cleaned mirror of the real corpora with a frozen offline enhancer (RE-USE \cite{fu2026rethinking}). The mixture is always built from the \emph{noisy} (acoustically real) target, while the enrollment is loaded from the \emph{cleaned} mirror. For the offline USEF-TFGridNet, the two views form a combined supervision target (noisy as the primary term, cleaned as a light auxiliary) that keeps the acoustic mapping faithful to the real target while improving perceptual quality; weights are given in Section~\ref{sec:training}. To ensure training stability, the online SwiftNet-Lookahead is supervised only with the noisy real target.

\section{Systems}
Both tracks use the data pipeline above, but differ in separator architecture and enrollment conditioning.

\subsection{Online track: SwiftNet-Lookahead}
For Track~1, \emph{SwiftNet-Lookahead} introduces bounded future context into a causal TSE model. We select SwiftNet because its causal iterative structure is compatible with the strict latency constraint. Its backbone is a weight-shared Swift-Net \cite{sang2025swiftnet} separator with twelve iterative updates, each combining multi-scale causal convolutions, frequency-axis bidirectional modeling, time-axis unidirectional modeling, and causally masked self-attention. A frozen ResNet-34 speaker encoder \cite{he2016resnet} from WeSpeaker \cite{wang2023wespeaker} provides the target representation, which is injected by feature-wise linear modulation (FiLM) \cite{perez2018film} only at the first iteration; subsequent updates refine the same conditioned representation. The STFT front end uses a 256-point window and 128-sample hop, adding only the analysis-synthesis delay.

A naive lookahead inside each shared update would accumulate: $n$ future frames per update would become $12n$ frames after twelve iterations. We instead insert a single \emph{LookaheadModule} between the audio bottleneck and the causal separator. It partitions time into non-overlapping chunks of length $S$ and applies a bidirectional LSTM independently within each chunk; the output is projected, layer-normalized, and added as a residual. Because no recurrent state crosses chunk boundaries, the maximum future context is
\begin{equation}
\text{lookahead}_{\max} = (S-1)\,H,
\label{eq:la}
\end{equation}
where $H$ is the STFT hop. The separator remains strictly causal, so no additional lookahead accumulates. We use a two-layer bidirectional LSTM with 128 hidden units per direction on the 256-channel bottleneck. With $S=11$ and $H=128$ samples (8 ms/frame), the maximum lookahead is $80$ ms.

Under the challenge latency definition, the algorithmic latency is the 8 ms STFT delay plus the 80 ms lookahead, yielding 88 ms; including the 8 ms hop buffer gives 96 ms total system latency. Streaming emits one hop once its bounded-lookahead chunk is available. A tail-perturbation probe reported effective lookaheads of 37.9, 75.0, and 22.6 ms at three positions, all below the structural bound in Eq.~\eqref{eq:la}; these spot checks supplement, but do not replace, the architectural latency bound.

\subsection{Offline track: USEF-TFGridNet}
For Track~2, we select USEF-TFGridNet \cite{zeng2025useftse,wang2023tfgridnet} for full-context modeling and frame-level enrollment conditioning, supported by offline inference. Rather than compressing the enrollment into a fixed-dimensional speaker embedding, the model keeps frame-level enrollment features and conditions the mixture through per-frequency cross-attention, with mixture frames as queries and effective enrollment frames as keys and values. Since padded enrollment frames are excluded from the keys and values, conditioning is invariant to in-batch padding. The target-aware representation is concatenated with the mixture representation, processed by six GridNetV2 blocks, and decoded by a transposed convolution followed by an inverse STFT. The model uses a 256-point window, a 128-sample hop, a shared encoder with $C=128$ channels, four attention heads, and GridNetV2 blocks with hidden dimension 256.

After extraction, we apply a validation-tuned magnitude-domain fusion designed to balance perceptual quality and speaker fidelity. Let $\mathbf{x}_r,\mathbf{x}_e\in\mathbb{R}^{N}$ denote the raw extractor output and the output after a frozen RE-USE \cite{fu2026rethinking} enhancer for the same sample. Enhancement can suppress residual noise but may also attenuate the target, so we retain target-speaker cues from the raw output. 

Specifically, we first align the RMS of the enhanced output to that of the original output:
\begin{equation}
\tilde{\mathbf{x}}_e = g\,\mathbf{x}_e,\quad
g=\operatorname{clip}\!\left(\frac{\operatorname{rms}(\mathbf{x}_r)}{\operatorname{rms}(\mathbf{x}_e)},\,0.1,\,10\right).
\end{equation}
Let $\mathbf{X}_r=\mathrm{STFT}(\mathbf{x}_r)$ and $\mathbf{X}_e=\mathrm{STFT}(\tilde{\mathbf{x}}_e)$, with magnitudes $\mathbf{M}_r=|\mathbf{X}_r|$ and $\mathbf{M}_e=|\mathbf{X}_e|$. We estimate frame-level speech activity from the enhanced magnitude:
\begin{equation}
e_t=\sqrt{\frac{1}{F}\sum_f \mathbf{M}_e(f,t)^2},\quad
a_t=\sigma\!\left(\frac{20\log_{10}\frac{e_t+\epsilon}{p_{20}+\epsilon}-\tau}{\Delta}\right),
\end{equation}
where $p_{20}$ is the 20th-percentile energy of $\{e_t\}$, $\tau=10$ dB is the activity threshold above this floor, $\Delta=6$ dB sets the sigmoid transition width, and $\epsilon$ is a small constant. We smooth $a_t$ with a 5-frame moving mean to obtain $\tilde{a}_t$, and compute the fused magnitude as
\begin{equation}
w_t=w-\rho \tilde{a}_t,\qquad
\mathbf{M}_f(f,t)=w_t\,\mathbf{M}_e(f,t)+(1-w_t)\mathbf{M}_r(f,t).
\label{eq:fusion}
\end{equation}
Here $w$ is the background-frame enhancement weight and $\rho$ reduces this weight on speech-active frames. We use validation-tuned values $w=0.9$ and $\rho=0.35$, selected on the same held-out validation meetings used for checkpoint selection (Section~\ref{sec:training}), so active frames retain more of the original TSE magnitude ($w-\rho=0.55$) while background frames remain closer to the enhanced magnitude. The output is
\begin{equation}
\hat{\mathbf{x}}=\mathrm{iSTFT}\!\left(\mathbf{M}_f \exp\left(j\angle \mathbf{X}_r\right)\right).
\end{equation}

\section{Training Setup}
\label{sec:training}
Both systems were trained with the same protocol: gradient clipping at norm 5.0 and a plateau learning-rate scheduler (decay factor 0.5) driven by the scale-invariant SNR on a validation set of held-out meetings with scenes disjoint from training. Checkpoint selection and early stopping both followed this validation metric, avoiding any tuning with leaderboard feedback.

SwiftNet-Lookahead minimized a plain SNR loss against the target speech, with its ResNet-34 speaker encoder frozen, and was trained on six GPUs with AdamW (initial learning rate $10^{-3}$, weight decay $0.1$). USEF-TFGridNet used a combined SNR loss weighting the noisy real target by $0.8$ and the cleaned mirror by $0.2$, and was optimized by Adam \cite{kingma2015adam} with an initial learning rate of $5\times10^{-4}$. All external corpora and frozen checkpoints used by the submission are named in Sections~II--III; no development or evaluation split is used for training, fine-tuning, or data augmentation.

\section{Results}
\textbf{Evaluation protocol.} Evaluation comprises EVAL-1 (1930 ``seen'' mixture--enrollment pairs drawn without overlap from development corpora) and EVAL-2 (3000 ``unseen'' real-world samples newly recorded across meeting rooms, caf\'es, homes, and vehicles with near-field, far-field, and mobile devices). Tables~\ref{tab:online}--\ref{tab:ablation_offline} report combined EVAL-1+EVAL-2 results. We follow the official scoring toolkit \cite{realtseChallenge2026}. Token error rate (TER) is computed as English WER or Chinese CER by comparing Zipformer-EN/ZH transcriptions with normalized reference transcripts. Timing F1 compares VAD-derived output activity with annotated target-speaker activity. Speaker similarity (SIM) is the cosine similarity between WeSpeaker \cite{wang2023wespeaker} embeddings of the extracted speech and enrollment. DNSMOS-P808 \cite{reddy2021dnsmos} estimates perceptual quality without a clean reference. Each metric is averaged over evaluation pairs; lower TER and higher values of the other metrics are better.

\textbf{Online track.} SwiftNet-Lookahead ranks third in the final verified Phase-1 ranking and exceeds both causal BSRNN baselines \cite{luo2023bandsplitrnn} on all four metrics (Table~\ref{tab:online}). Table~\ref{tab:ablation_online} compares training data and enrollment encoders on EVAL-1+EVAL-2. Mixed synthetic and real training gives the best TER, Timing F1, and SIM. Synthetic-only training yields TER $0.703$ and DNSMOS-P808 $2.173$; real-only training gives the highest DNSMOS-P808 but lower intelligibility, Timing F1, and SIM than mixed training. These comparisons support combining real acoustics with broader synthetic variation. ResNet-34 gives the best TER and DNSMOS-P808 among enrollment encoders. TF-MAP/contextual embedding \cite{zhang2025multilevel} gives slightly higher Timing F1 and SIM but lower DNSMOS-P808 ($2.494$); ECAPA-TDNN \cite{desplanques2020ecapa} has comparable TER but weaker remaining metrics. We therefore use mixed training and ResNet-34 conditioning.

\begin{table}[htbp]
\caption{Online track (Track 1).}
\vspace{-10pt}
\begin{center}
\footnotesize
\setlength{\tabcolsep}{2.5pt}
\resizebox{\columnwidth}{!}{%
\begin{tabular}{@{}lcccc@{}}
\toprule
\textbf{System} & \textbf{TER}$\downarrow$ & \textbf{Timing F1}$\uparrow$ & \textbf{SIM}$\uparrow$ & \textbf{DNSMOS-P808}$\uparrow$ \\
\midrule
SwiftNet-Lookahead (ours) & \textbf{0.590} & \textbf{0.848} & \textbf{0.480} & \textbf{3.111} \\
BSRNN-TFMAP & 0.712 & 0.835 & 0.456 & 2.785 \\
BSRNN-EMB & 0.734 & 0.821 & 0.422 & 2.737 \\
\bottomrule
\end{tabular}
}
\label{tab:online}
\end{center}
\vspace{-10pt}
\end{table}

\begin{table}[htbp]
\caption{SwiftNet-Lookahead ablations (Track 1).}
\vspace{-10pt}
\begin{center}
\footnotesize
\setlength{\tabcolsep}{2pt}
\resizebox{\columnwidth}{!}{%
\begin{tabular}{@{}lcccc@{}}
\toprule
\textbf{Configuration} & \textbf{TER}$\downarrow$ & \textbf{Timing F1}$\uparrow$ & \textbf{SIM}$\uparrow$ & \textbf{DNSMOS-P808}$\uparrow$ \\
\midrule
\multicolumn{5}{@{}l}{Training-data composition} \\
Mixed synthetic + real (default) & \textbf{0.590} & \textbf{0.848} & \textbf{0.480} & 3.111 \\
Synthetic data only & 0.703 & 0.840 & 0.454 & 2.173 \\
Real data only & 0.606 & 0.842 & 0.450 & \textbf{3.212} \\
\midrule
\multicolumn{5}{@{}l}{Enrollment audio encoder} \\
ResNet-34 (default) & \textbf{0.590} & 0.848 & 0.480 & \textbf{3.111} \\
ECAPA-TDNN & 0.593 & 0.829 & 0.464 & 3.102 \\
TF-MAP/contextual embedding & 0.621 & \textbf{0.849} & \textbf{0.483} & 2.494 \\
\bottomrule
\end{tabular}
}
\label{tab:ablation_online}
\end{center}
\end{table}

\textbf{Offline track.} USEF-TFGridNet ranks fifth in the final verified Phase-1 ranking and exceeds both baselines (Table~\ref{tab:offline}), with TER $0.538$ and DNSMOS-P808 $3.256$. Magnitude fusion is designed to balance perceptual quality and speaker fidelity, motivated by the perception-distortion trade-off \cite{blau2018perception}. In the ablations (Table~\ref{tab:ablation_offline}), frame-level enrollment cross-attention gives the best TER and DNSMOS-P808; compressed embeddings have similar Timing F1/SIM but lower perceptual quality. The $0.8{:}0.2$ supervision weight gives the best values on all four metrics within its block. These ablations do not isolate lookahead or magnitude-fusion gains.

\begin{table}[htbp]
\caption{Offline track (Track 2).}
\vspace{-10pt}
\begin{center}
\footnotesize
\setlength{\tabcolsep}{2.5pt}
\resizebox{\columnwidth}{!}{%
\begin{tabular}{@{}lcccc@{}}
\toprule
\textbf{System} & \textbf{TER}$\downarrow$ & \textbf{Timing F1}$\uparrow$ & \textbf{SIM}$\uparrow$ & \textbf{DNSMOS-P808}$\uparrow$ \\
\midrule
USEF-TFGridNet (ours) & \textbf{0.538} & \textbf{0.851} & \textbf{0.472} & \textbf{3.256} \\
BSRNN-EMB (baseline) & 0.739 & 0.829 & 0.418 & 2.874 \\
BSRNN-TFMAP (baseline) & 0.745 & 0.829 & 0.444 & 2.755 \\
\bottomrule
\end{tabular}
}
\label{tab:offline}
\end{center}
\vspace{-10pt}
\end{table}

\begin{table}[htbp]
\caption{USEF-TFGridNet ablations (Track 2).}
\vspace{-10pt}
\begin{center}
\footnotesize
\setlength{\tabcolsep}{2pt}
\resizebox{\columnwidth}{!}{%
\begin{tabular}{@{}lcccc@{}}
\toprule
\textbf{Configuration} & \textbf{TER}$\downarrow$ & \textbf{Timing F1}$\uparrow$ & \textbf{SIM}$\uparrow$ & \textbf{DNSMOS-P808}$\uparrow$ \\
\midrule
\multicolumn{5}{@{}l}{Enrollment audio encoder} \\
USEF (default) & \textbf{0.538} & 0.851 & 0.472 & \textbf{3.256} \\
ResNet-34 & 0.585 & \textbf{0.853} & 0.475 & 2.494 \\
ECAPA-TDNN & 0.576 & 0.850 & \textbf{0.477} & 2.947 \\
TF-MAP/contextual embedding & 0.618 & 0.845 & 0.471 & 2.802 \\
\midrule
\multicolumn{5}{@{}l}{\emph{Combined supervision weight (noisy real target:cleaned mirror target)}} \\
$0:1$ & 0.598 & 0.844 & 0.455 & 2.977 \\
$0.5:0.5$ & 0.585 & 0.841 & 0.451 & 3.247 \\
$0.8:0.2$ (default) & \textbf{0.538} & \textbf{0.851} & \textbf{0.472} & \textbf{3.256} \\
$1:0$ & 0.541 & 0.840 & 0.453 & 2.948 \\
\bottomrule
\end{tabular}
}
\label{tab:ablation_offline}
\end{center}
\vspace{-10pt}
\end{table}

\section{Conclusion}
This paper presents the SonicAGI systems for the online and offline tracks of the REAL-TSE Challenge. The results suggest that real conversational TSE depends not only on separator design, but also on training-data realism and target construction. Our two-route data pipeline combines controllable synthetic mixtures with real meeting overlap, and uses a cleaned mirror target to reduce the effect of residual background noise in real recordings. SwiftNet-Lookahead confines future context to a single bounded injection, achieving third place in Track~1 with 96 ms total system latency. USEF-TFGridNet combines frame-level enrollment cross-attention with magnitude-domain fusion, ranking fifth in Track~2 and outperforming the challenge baselines. The trade-off between speaker similarity and perceptual quality remains an important design consideration; future work will explore similarity-aware training or fusion objectives and extend bounded lookahead to stronger cross-attention extractors.

\clearpage

\section*{AI-Generated Content Disclosure}
ChatGPT was used to assist with language editing, grammar refinement, and concision of the manuscript text. The authors reviewed and revised all AI-assisted edits.

\bibliographystyle{IEEEtran}
\bibliography{real_tse_refs}

\end{document}